\documentclass{mem}
\usepackage{natbib}\usepackage{txfonts}\usepackage{balance}
\usepackage{graphicx}
\usepackage[a4paper,breaklinks,dvipdfm]{hyperref}
\idline{0}{0}
\begin{document}
\def\kms{$\mathrm {km s}^{-1}$}

\title{
Astrometry of brown dwarfs with Gaia
}

  \subtitle{}

\author{
J.H.J. \,de Bruijne%\inst{1} 
       }

% \offprints{J.H.J. de Bruijne}

\institute{
Scientific Support Office in the Directorate of Science and Robotic Exploration of the European Space Agency, Postbus 299, 2200AG, Noordwijk, The Netherlands\break
\email{jos.de.bruijne@esa.int}
}

\authorrunning{De Bruijne}

\titlerunning{Astrometry of brown dwarfs with Gaia}

\abstract{
Europe's Gaia spacecraft will soon embark on its five-year mission to measure the absolute parallaxes of the complete sample of $1,000$ million objects down to $20$ mag. It is expected that thousands of nearby brown dwarfs will have their astrometry determined with sub-milli-arcsecond standard errors. Although this level of accuracy is comparable to the standard errors of the relative parallaxes that are now routinely obtained from the ground for selected, individual objects, the absolute nature of Gaia's astrometry, combined with the sample increase from one hundred to several thousand sub-stellar objects with known distances, ensures the uniqueness of Gaia's legacy in brown-dwarf science for the coming decade(s). We shortly explore the gain in brown-dwarf science that could be achieved by lowering Gaia's faint-end limit from 20 to 21 mag and conclude that two spectral-type sub-classes could be gained in combination with a fourfold increase in the solar-neighbourhood-volume sampled by Gaia and hence in the number of brown dwarfs in the Gaia~Catalogue.
}
\maketitle{}

\section{Introduction}

\begin{figure*}[ht!]
\center{\resizebox{0.99\hsize}{!}{\includegraphics[clip=true,angle=270]{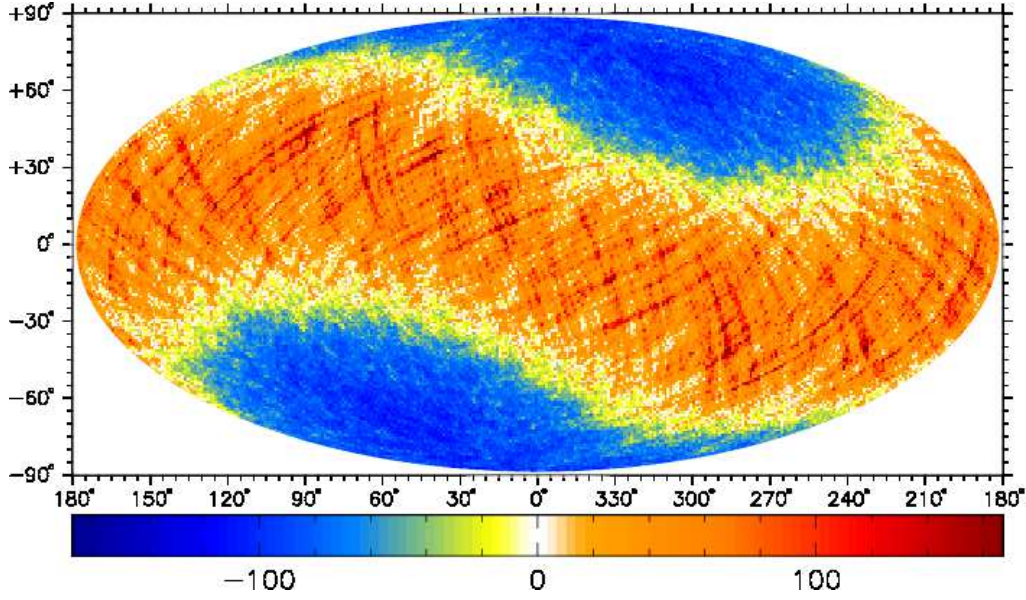}}}
\caption{\footnotesize
Variation over the sky, in equatorial coordinates, of the difference between the sky-average, end-of-mission parallax standard error at $G=20$~mag ($\sigma_\pi = 332~\mu$as, colour coded in white) and the local, end-of-mission parallax standard error, in units of $\mu$as, from \url{http://www.cosmos.esa.int/web/gaia/science-performance}. The variation is caused by Gaia's scanning law. The red, curved band follows the ecliptic plane and the blue areas denote the ecliptic-pole regions.}
\label{fig:sky_map}
\end{figure*}

\begin{table}[ht!]
\caption{Sky-averaged Gaia astrometric standard errors between $G = 20$ and $21$~mag for three assumed detection probabilities. The [min--max] range for the parallax error denotes the variation over the sky. Even with modest detection percentages ($p_{\rm det} = 50$\%), Gaia can deliver sub-mas astrometry at $G = 21$~mag.}
\label{tab:faint_end}
\begin{center}
\begin{tabular}{cccc}
\\[-25pt]
\hline
\\[-4pt]
$G$ & $\sigma_\pi$ [min -- max] & $\sigma_0$ & $\sigma_\mu$\\
mag & $\mu$as                   & $\mu$as    & $\mu$as~year$^{-1}$\\
\\[-4pt]
\hline
\\[-4pt]
\multicolumn{4}{c}{Detection probability $p_{\rm det} = 100$\%}\\
\\[-8pt]
20.0 & 332~ [233 -–  \phantom{1}384] & 247 & 175\\
20.5 & 466~ [326 -–  \phantom{1}539] & 346 & 245\\
21.0 & 670~ [469 -–  \phantom{1}775] & 498 & 353\\
\\
\multicolumn{4}{c}{Detection probability $p_{\rm det} = 80$\%}\\
\\[-8pt]
20.0 & 372~ [260 -–  \phantom{1}430] & 276 & 195\\
20.5 & 521~ [365 -–  \phantom{1}602] & 387 & 274\\
21.0 & 749~ [525 -–  \phantom{1}866] & 557 & 394\\
\\
\multicolumn{4}{c}{Detection probability $p_{\rm det} = 50$\%}\\
\\[-8pt]
20.0 & 470~ [329 -–  \phantom{1}543] & 349 & 247\\
20.5 & 659~ [461 -–  \phantom{1}762] & 489 & 347\\
21.0 & 948~ [664 -– 1096] & 704 & 499\\
\\[-4pt]
\hline
\end{tabular}
\end{center}
\end{table}

Gaia is the current astrometry mission of the European Space Agency (ESA), following up on the success of the Hipparcos mission. With a focal plane containing $106$ CCD detectors, Gaia is about to start surveying the entire sky and repeatedly observe the brightest $1,000$ million objects, down to $20^{\rm th}$ magnitude, during its five-year lifetime. Gaia's science data comprises absolute astrometry, broad-band photometry, and low-resolution spectro-photometry. Spectroscopic data with a resolving power of $11,500$ will be obtained for the brightest $150$ million sources, down to $17^{\rm th}$ magnitude. The thermo-mechanical stability of the spacecraft, combined with the selection of the L2 Lissajous point of the Sun-Earth/Moon system for operations, allows parallaxes to be measured with standard errors less than $10$~micro-arcsecond ($\mu$as) for stars brighter than $12^{\rm th}$ magnitude, $25~\mu$as for stars at $15^{\rm th}$ magnitude, and $300~\mu$as at magnitude $20$. Photometric standard errors are in the milli-magnitude regime. The spectroscopic data allows the measurement of radial velocities with errors of $15$~km~s$^{-1}$ at magnitude $17$. Gaia's primary science goal is to unravel the kinematical, dynamical, and chemical structure and evolution of the Milky Way. In addition, Gaia's data will revolutionise many other areas of science, e.g., stellar physics, solar-system bodies, fundamental physics, exo-planets, and -- last but not least -- brown dwarfs. The Gaia spacecraft has been launched on 19 December $2013$ and will start its science mission in the summer of $2014$. The science community in Europe, organised in the Data Processing and Analysis Consortium (DPAC), is responsible for the processing of the data. The first intermediate data is expected to be released some two years after launch while the final catalogue is expected around $2022$. ESA's community web portal \url{http://www.cosmos.esa.int/gaia} provides more information on the Gaia mission, including~bibliographies.

\section{Gaia parallaxes}

Gaia's science performance has been studied by \cite{2012Ap&SS.341...31D} and is also presented on \url{http://www.cosmos.esa.int/web/gaia/science-performance}. In short, for a given $V$ magnitude and $V-I$ colour index, the end-of-mission parallax standard error, $\sigma_\pi$ in units of $\mu$as, averaged over the sky, can be calculated as:
\begin{eqnarray}
\sigma_\pi &=& [9.3 + 658.1 \cdot z + 4.568 \cdot z^2]^{1/2} \cdot \nonumber\\
           & & \cdot [0.986 + (1 - 0.986)\cdot (V-I)],
\end{eqnarray}
where:
\begin{equation}
z = {\rm MAX}[10^{0.4\cdot(12 - 15)}, 10^{0.4\cdot(G - 15)}],
\end{equation}
is a flux-based, auxiliary quantity, taking saturation at $G = 12$~mag into account, and \cite[following][but updated with the most recent coefficients]{2010A&A...523A..48J}:
\begin{eqnarray}
G &=& V - 0.0208 - 0.1004 \cdot (V-I) + \nonumber\\
  & & - 0.1593 \cdot (V-I)^2 + 0.0083 \cdot (V-I)^3
\end{eqnarray}
denotes the broad-band Gaia magnitude. For a five-year Gaia mission, the sky-averaged position and proper-motion standard errors, $\sigma_0$ and $\sigma_\mu$ in units of $\mu$as and $\mu$as~year$^{-1}$, respectively, are:
\begin{eqnarray}
\sigma_0   &=& 0.743 \cdot \sigma_\pi;\\
\sigma_\mu &=& 0.526 \cdot \sigma_\pi.
\end{eqnarray}
For example, at $G=20$~mag, one finds $\sigma_\pi = 332~\mu$as. Figure~\ref{fig:sky_map} shows the variation of $\sigma_\pi$ over the sky as induced by Gaia's scanning law.

\section{Gaia's faint-end limit}

\begin{figure*}[ht!]
\center{\resizebox{0.60\hsize}{!}{\includegraphics[clip=true]{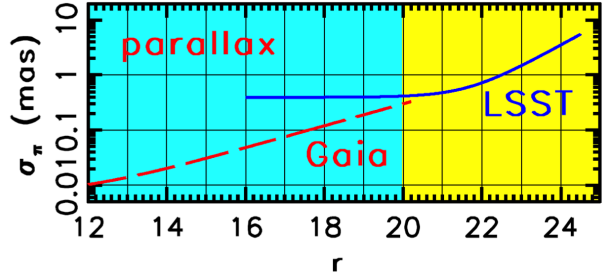}}}
\caption{\footnotesize
End-of-mission parallax standard error, in milli-arcsecond (mas), for Gaia (dashed, red line) and the Large Synoptic Survey Telescope (blue; LSST), as function of $r$ magnitude, from \cite{2012IAUS..282...33E}. Following \cite{2010A&A...523A..48J}, but updated with the most recent coefficients, we have $G - r = - 0.0501 + 0.1598 \cdot (r-i) - 0.5613 \cdot (r-i)^2 + 0.0628 \cdot (r-i)^3$. Figure courtesy Laurent Eyer and \u{Z}eljko Ivezi\'{c}.}
\label{fig:LSST}
\end{figure*}

Gaia’s default faint-star limit is $G = 20$~mag. This value, however, is a configurable parameter in the on-board detection software. Investigations by \cite{2013A&A...000..000D} have shown that, in principle, Gaia's intrinsic detection probability is very close to $100$\% all the way down to $G=21$~mag. Simply ignoring the practical complications and programmatic implications, a natural question to ask is whether it is scientifically interesting to lower Gaia's survey limit from $G = 20$ to 21~mag. The extra science that is available in this magnitude range comprises proper motions of distant, faint halo stars \citep[e.g.,][]{2012sf2a.conf..113R}, astrometry of white dwarfs in the solar neighbourhood \citep[e.g.,][]{2014arXiv1403.6045C}, and solar-neighbourhood brown-dwarf science such as the sub-stellar mass function, three-dimensional modelling of cool atmospheres, etc. \citep[e.g., ][]{2002EAS.....2..199H,2013A&A...550A..44S}. Additional science cases which would also benefit from going deeper include ultra-faint dwarf galaxies, asteroids, and so-called global parameters in the astrometric global iterative solution for which quasars are used \citep[e.g., reference-frame parameters, the energy flux of primordial gravity waves, or the acceleration of the solar-system barycentre; e.g.,][]{2012IAUJD...7P...7B}. 

\begin{figure*}[ht!]
\resizebox{1.00\hsize}{!}{\includegraphics[clip=true]{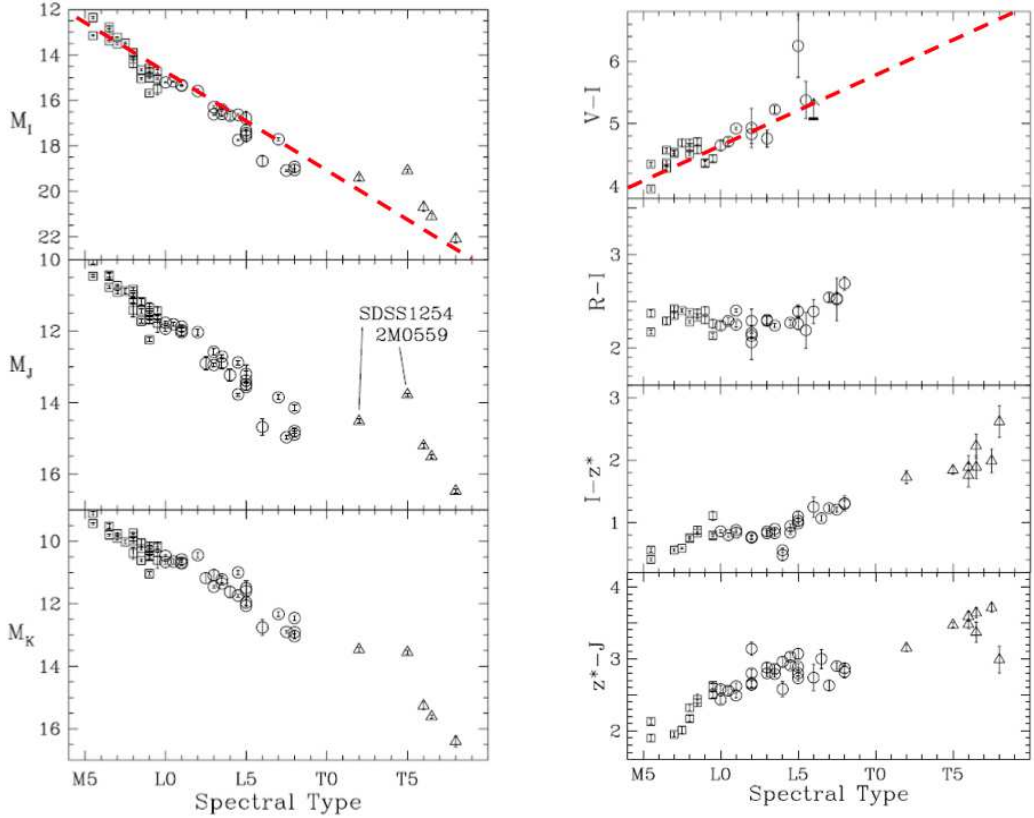}}
\caption{\footnotesize
Colour, absolute-magnitude, and spectral-type relations for late-type objects, from \cite{2002AJ....124.1170D}. The dashed red lines indicate the linear relations -- ``fit by eye'' -- used in this work (see Table~\ref{tab:astrometry}).}
\label{fig:cal}
\end{figure*}

Gaia's capabilities to deliver all-sky, absolute astrometry are unique and a successor mission is not likely to be launched before the mid 2030's. Clearly, one can do (relative) astrometry from the ground and one particularly interesting project currently in development is the Large Synoptic Survey Telescope (LSST, \url{http://www.lsst.org/}), which will provide six-band $(u,g,r,i,z,y)$ photometry for some $10$~billion stars down to $24^{\rm th}$~mag in $20,000$~deg$^{2}$ observed twice per week during $10$~years. The predictions of the precision of the astrometry that LSST will provide in the Gaia reference frame are shown in Figure~\ref{fig:LSST}, which suggests that there is a gap between Gaia's nominal faint limit at $20$~mag and LSST's bright limit around $21$~mag. This gap can, in principle, be closed by Gaia, after which LSST would provide a near-perfect extension of Gaia by $\sim$$4$ magnitudes.

\section{Gaia astrometry for brown dwarfs}

\begin{table*}[ht!]
\caption{Brown-dwarf astrometry with Gaia.
(1) Spectral type.
(2) Absolute $I$-band magnitude (see Figure~\ref{fig:cal}).
(3) $V-I$ colour index (see Figure~\ref{fig:cal}).
(4) Absolute $V$-band magnitude (see Figure~\ref{fig:cal}).
(5) $V$ magnitude at $30$~pc distance.
(6) $G$ magnitude at $30$~pc distance. With a limiting magnitude of $G = 20$~mag, L4 brown dwarfs can be seen out to $30$~pc. At $G = 21$~mag, this would improve to L6 dwarfs at $30$~pc.
(7) Distance limit, in pc, corresponding to $G = 20.0$~mag.
(8) Relative parallax error at that distance (assuming $p_{\rm det} = 100$\%). A 1\% relative parallax error is reached for an L4 dwarf at 30~pc.
(9) Distance limit, in pc, corresponding to $G = 20.5$~mag.
(10) Relative parallax error at that distance (assuming $p_{\rm det} = 80$\%). A 1\% relative parallax error is reached for an L7 dwarf at 20~pc.
(11) Distance limit, in pc, corresponding to $G = 21.0$~mag.
(12) Relative parallax error at that distance (assuming $p_{\rm det} = 50$\%). A 1\% relative parallax error is reached for a T1 dwarf at 10~pc.
}
\label{tab:astrometry}
\begin{center}
\begin{tabular}{ccccccccccccc}
\\[-32pt]
\hline
\\[-4pt]
(1) & (2) & (3) & (4) & (5) & (6) & (7) & (8) & (9) & (10) & (11) & (12)\\
SpT & $M_I$ & $V-I$ & $M_V$ & $V$  & $G$  & $d$ & $\sigma_\pi / \pi$ & $d$ & $\sigma_\pi / \pi$ & $d$ & $\sigma_\pi / \pi$\\
    & mag   & mag   & mag   & mag         & mag         & pc &  \%                 & pc  & \%                & pc  & \%                \\
    &       &       &       & $@ 30$~pc            & $@ 30$~pc            & \multicolumn{2}{c}{$G=20.0$~mag} & \multicolumn{2}{c}{$G=20.5$~mag} & \multicolumn{2}{c}{$G=21.0$~mag}\\
    &       &       &       &             &             & \multicolumn{2}{c}{$p_{\rm det} = 100$\%} & \multicolumn{2}{c}{$p_{\rm det} = 80$\%} & \multicolumn{2}{c}{$p_{\rm det} = 50$\%}\\
\\[-4pt]
\hline
\\[-4pt]
M5	&12.00	&4.00	&16.00	&18.39	&15.95	&194	&6.45	&244	&12.73	&308	&29.16\\
M6	&12.46	&4.12	&16.58	&18.96	&16.41	&157	&5.22	&198	&10.30	&249	&23.59\\
M7	&12.92	&4.23	&17.15	&19.54	&16.86	&127	&4.23	&160	&8.33	&201	&19.10\\
M8	&13.38	&4.35	&17.73	&20.11	&17.32	&103	&3.42	&130	&6.75	&163	&15.47\\
M9	&13.83	&4.47	&18.30	&20.69	&17.78	&83	&2.77	&105	&5.47	&132	&12.54\\
L0	&14.29	&4.58	&18.88	&21.26	&18.23	&68	&2.25	&85	&4.44	&107	&10.17\\
L1	&14.75	&4.70	&19.45	&21.84	&18.69	&55	&1.83	&69	&3.60	&87	&8.26\\
L2	&15.21	&4.82	&20.03	&22.41	&19.14	&45	&1.48	&56	&2.93	&71	&6.70\\
L3	&15.67	&4.93	&20.60	&22.99	&19.59	&36	&1.20	&46	&2.38	&57	&5.45\\
L4	&16.13	&5.05	&21.18	&23.56	&20.04	&29	&0.98	&37	&1.93	&47	&4.43\\
L5	&16.58	&5.17	&21.75	&24.14	&20.49	&24	&0.80	&30	&1.57	&38	&3.60\\
L6	&17.04	&5.28	&22.33	&24.71	&20.94	&19	&0.65	&25	&1.28	&31	&2.93\\
L7	&17.50	&5.40	&22.90	&25.29	&21.38	&16	&0.53	&20	&1.04	&25	&2.38\\
L8	&17.96	&5.52	&23.48	&25.86	&21.83	&13	&0.43	&16	&0.85	&20	&1.94\\
L9	&18.42	&5.63	&24.05	&26.44	&22.28	&11	&0.35	&13	&0.69	&17	&1.58\\
T0	&18.88	&5.75	&24.63	&27.01	&22.72	&9	&0.28	&11	&0.56	&14	&1.29\\
T1	&19.33	&5.87	&25.20	&27.59	&23.17	&7	&0.23	&9	&0.46	&11	&1.05\\
T2	&19.79	&5.98	&25.78	&28.16	&23.61	&6	&0.19	&7	&0.37	&9	&0.85\\
T3	&20.25	&6.10	&26.35	&28.74	&24.06	&5	&0.15	&6	&0.30	&7	&0.70\\
T4	&20.71	&6.22	&26.93	&29.31	&24.50	&4	&0.13	&5	&0.25	&6	&0.57\\
T5	&21.17	&6.33	&27.50	&29.89	&24.95	&3	&0.10	&4	&0.20	&5	&0.46\\
T6	&21.63	&6.45	&28.08	&30.46	&25.39	&3	&0.08	&3	&0.16	&4	&0.38\\
T7	&22.08	&6.57	&28.65	&31.04	&25.84	&2	&0.07	&3	&0.13	&3	&0.31\\
T8	&22.54	&6.68	&29.23	&31.61	&26.28	&2	&0.06	&2	&0.11	&3	&0.25\\
T9	&23.00	&6.80	&29.80	&32.19	&26.73	&1	&0.05	&2	&0.09	&2	&0.20\\
\\[-4pt]
\hline
\end{tabular}
\end{center}
\end{table*}

So far, about $100$ solar-neighbourhood brown dwarfs have had their relative parallaxes measured from the ground with percent-level precisions and (sub-)mas-level standard errors \citep[e.g.,][see DwarfArchives.org for a compilation]{2012ApJ...752...56F,2013MNRAS.433.2054S,2013AJ....146..161M,2013A&A...560A..52M,2014PASP..126...15W}. In the area of parallax standard errors, Gaia will perform better but not spectacularly: most brown dwarfs will be faint and hence will have Gaia parallax standard errors in the range of a few hundred $\mu$as (Table~\ref{tab:faint_end}). One should, however, recall that Gaia's astrometry is absolute and not relative. In the area of sample size, estimations are that a few thousand brown dwarfs with $G < 20$~mag will be accessible to Gaia, which constitutes a significant increase in number from the current $\sim$$100$ dwarfs with parallaxes; in addition, Gaia's all-sky brown-dwarf sample will not be hampered by (kinematic and/or spatial) selection effects such as the galactic-plane avoidance zone in the current sample of known brown dwarfs \citep[see Figure~1 in][]{2009MmSAI..80..674S}. If Gaia's faint limit would be extended by $0.5$~mag, the local galactic volume sampled by Gaia would increase by a factor two; if the magnitude limit would be lowered by $1.0$~mag, the volume and sample-size increase would be a factor~four.

Table~\ref{tab:astrometry} shows what Gaia can achieve in terms of parallax standard errors as function of spectral type. There are three ways in which one can quantify the gain of going one magnitude deeper:
\begin{enumerate}
\item One magnitude deeper gains two brown-dwarf spectral-type sub-classes at a given distance:
\begin{enumerate}
\item $G = 20.0$~mag allows, for instance, reaching L4 brown dwarfs at $30$~pc (or L6 ones at $20$~pc);
\item $G = 20.5$~mag allows, for instance, reaching L5 brown dwarfs at $30$~pc (or L7 ones at $20$~pc);
\item $G = 21.0$~mag allows, for instance, reaching L6 brown dwarfs at $30$~pc (or L8 ones at $20$~pc);
\end{enumerate}
\item One magnitude deeper increases the distance range for a given spectral type by $60$\% and the volume sampled by $300$\%:
\begin{enumerate}
\item $G = 20.0$~mag allows reaching L5 brown dwarfs at $24$~pc with $0.8$\% relative parallax error (for $p_{\rm det} = 100$\%);
\item $G = 20.5$~mag allows reaching L5 brown dwarfs at $30$~pc with $1.6$\% relative parallax error (for $p_{\rm det} = 80$\%);
\item $G = 21.0$~mag allows reaching L5 brown dwarfs at $38$~pc with $3.6$\% relative parallax error (for $p_{\rm det} = 50$\%);
\end{enumerate}
\item One magnitude deeper increases the distance at which a $1$\% relative parallax error is reached and/or moves the spectral classification to later types:
\begin{enumerate}
\item $G = 20.0$~mag allows reaching a $1$\% relative parallax error for an L4 dwarf at $30$~pc (assuming $p_{\rm det} = 100$\%);
\item $G = 20.5$~mag allows reaching a $1$\% relative parallax error for an L7 dwarf at $20$~pc (assuming $p_{\rm det} = 80$\%);
\item $G = 21.0$~mag allows reaching a $1$\% relative parallax error for a  T1 dwarf at $10$~pc (assuming $p_{\rm det} = 50$\%).
\end{enumerate}
\end{enumerate}

\section{Conclusions}

With a nominal survey limit at $G = 20$~mag, Gaia will make a significant contribution to brown-dwarf science by delivering sub-milli-arcsecond absolute astrometry of an unbiased, all-sky sample of thousands of brown dwarfs in the solar neighbourhood. By lowering the limit to $G = 21$~mag, this sample can, in principle, i.e., when ignoring programmatic limitations, be increased in number by a factor four.

\begin{acknowledgements}
It is a pleasure to thank the local organising committee, and Ricky Smart in particular, for an enjoyable workshop. This research has benefited from the M, L, T, and Y dwarf compendium compiled by Chris Gelino, Davy Kirkpatrick, Mike Cushing, David Kinder, and Adam Burgasser and housed at DwarfArchives.org.
\end{acknowledgements}

\bibliographystyle{aa}
\bibliography{josdebruijne}

\end{document}